\documentstyle[12pt,psfig]{article}

\begin{document}
\baselineskip 0.7cm

\newcommand{\gsim}{ \mathop{}_{\textstyle \sim}^{\textstyle >} }
\newcommand{\lsim}{ \mathop{}_{\textstyle \sim}^{\textstyle <} }
\newcommand{\vev}[1]{ \left\langle {#1} \right\rangle }
\newcommand{\lsp}{ \left ( }
\newcommand{\rsp}{ \right ) }
\newcommand{\lmp}{ \left \{ }
\newcommand{\rmp}{ \right \} }
\newcommand{\llp}{ \left [ }
\newcommand{\rlp}{ \right ] }
\newcommand{\labs}{ \left | }
\newcommand{\rabs}{ \right | }
\newcommand{\EV} { {\rm eV} }
\newcommand{\KEV}{ {\rm keV} }
\newcommand{\MEV}{ {\rm MeV} }
\newcommand{\GEV}{ {\rm GeV} }
\newcommand{\TEV}{ {\rm TeV} }
\newcommand{\YR}{ {\rm yr} }
\newcommand{\mgut}{M_{GUT}}
\newcommand{\mint}{M_{I}}
\newcommand{\mgra}{M_{3/2}}
\newcommand{\mll}{m_{\tilde{l}L}^{2}}
\newcommand{\mdr}{m_{\tilde{d}R}^{2}}
\newcommand{\mllXX}[1]{m_{\tilde{l}L , {#1}}^{2}}
\newcommand{\mdrXX}[1]{m_{\tilde{d}R , {#1}}^{2}}
\newcommand{\mgy}{m_{G1}}
\newcommand{\mgl}{m_{G2}}
\newcommand{\mgc}{m_{G3}}
\newcommand{\nuR}{\nu_{R}}
\newcommand{\slL}{\tilde{l}_{L}}
\newcommand{\slLi}{\tilde{l}_{Li}}
\newcommand{\sdR}{\tilde{d}_{R}}
\newcommand{\sdRi}{\tilde{d}_{Ri}}
\newcommand{\e}{{\rm e}}
\newcommand{\bsub}{\begin{subequations}}
\newcommand{\esub}{\end{subequations}}
\renewcommand{\thefootnote}{\fnsymbol{footnote}}
\setcounter{footnote}{1}

\makeatletter
%
%
%
%
%
\newtoks\@stequation

\def\subequations{\refstepcounter{equation}%
  \edef\@savedequation{\the\c@equation}%
  \@stequation=\expandafter{\theequation}
  \edef\@savedtheequation{\the\@stequation}
  \edef\oldtheequation{\theequation}%
  \setcounter{equation}{0}%
  \def\theequation{\oldtheequation\alph{equation}}}

\def\endsubequations{%
  \ifnum\c@equation < 2 \@warning{Only \the\c@equation\space subequation
    used in equation \@savedequation}\fi
  \setcounter{equation}{\@savedequation}%
  \@stequation=\expandafter{\@savedtheequation}%
  \edef\theequation{\the\@stequation}%
  \global\@ignoretrue}


\def\eqnarray{\stepcounter{equation}\let\@currentlabel\theequation
\global\@eqnswtrue\m@th
\global\@eqcnt\z@\tabskip\@centering\let\\\@eqncr
$$\halign to\displaywidth\bgroup\@eqnsel\hskip\@centering
     $\displaystyle\tabskip\z@{##}$&\global\@eqcnt\@ne
      \hfil$\;{##}\;$\hfil
     &\global\@eqcnt\tw@ $\displaystyle\tabskip\z@{##}$\hfil
   \tabskip\@centering&\llap{##}\tabskip\z@\cr}

\makeatother


\begin{titlepage}

\begin{flushright}
UT-973
\end{flushright}

\vskip 0.35cm
\begin{center}
{\large \bf Production and Detection of Black Holes \\
at Neutrino Array}
\vskip 1.2cm
Yosuke Uehara

\vskip 0.4cm

{\it Department of Physics, University of Tokyo, Tokyo 113-0033 Japan}

\vskip 1.5cm

\abstract{We consider the production of black holes caused by the
 collision between high-energy cosmic neutrinos and nuclei contained
in detectors. If the fundamental scale $M_{*}$ is ${\rm O}(\TEV)$,
as some higher-dimensional theories suggest, ICECUBE detector
may observe about $10^{4} \sim 10^{2}$ black holes per year.}

\end{center}
\end{titlepage}

\renewcommand{\thefootnote}{\arabic{footnote}}
\setcounter{footnote}{0}

%
%
%
%

Collision of particles can produce black holes if their
energy is higher than the Planck scale, $M_{pl}$. It was generally
thought that the Planck scale is so large, $M_{pl} = G^{-1/2} = 1.2 \times
10^{19} \GEV$, that our experiments could not create black holes.

However, if our world is higher-dimensional and the fundamental scale
is TeV scale \cite{LARGEEXTRADIMENSION,EXTRADIMENSIONSTRING,SMALLEXTRADIMENSION}, we can access black holes by real experiments. 
The production of black holes at the Large Hadron Collider(LHC) is considered
\cite{FIRSTBH,BHFACTORY,BHLHC,QUASISTABLEBH,BHLARGEEXTRADIMENSION}.
Black holes also can be created by high-energy cosmic rays
\cite{BHAIRSHOWER,CRLARGEEXTRADIMENSION,BHCR}. The detection
of black holes enables us to investigate the physics of
quantum gravity directly, and its detailed study must be done.

In this paper, we consider the detection of black holes
at neutrino array like ICECUBE. Black holes are produced by the collision
between high-energy neutrino cosmic rays and nuclei contained in the
ice of ICECUBE detector. The huge amount of ice in ICECUBE 
enables us to observe many black holes 
if the fundamental scale is ${\rm O}(\TEV)$.
Once the black hole is produced, its evaporation leads to a very clean
signal and it cannot escape the detection.

The Schwarzshild radius $R_{S}$ of a $(4+n)$-dimensional
black hole is \cite{BHCROSSSECTION}:
\begin{eqnarray}
R_{S} = \frac{1}{\sqrt{\pi} M_{*}} [ \frac{M_{BH}}{M_{*}} (\frac{8 \Gamma(\frac{n+3}{2})}{n+2})]^{\frac{1}{n+1}}.
\end{eqnarray}

Here $M_{*}$ is the fundamental scale of higher-dimensional world,
and $M_{BH}$ is the mass of black holes. We only consider semiclassical
black holes. This means we do not use the quantum gravity, but use the general
relativity only. Then, the production cross section of black holes is 
estimated as
\begin{eqnarray}
\sigma(M_{BH}) \sim \pi R_{S}^{2} = \frac{1}{M_{*}^{2}}[\frac{M_{BH}}{M_{*}}(\frac{8 \Gamma(\frac{n+3}{2})}{n+2})]^{\frac{2}{n+1}}.
\end{eqnarray}
This is valid if the mass of black hole
$M_{BH}$ is much larger than the fundamental scale $M_{*}$.
\footnote{Although, there are some arguments which do not support this
naive cross section \cite{BHSUPPRESSION1,BHSUPPRESSION2}.}

We consider the collision of high-energy neutrinos off nuclei.
Typically high-energy neutrinos are produced by the decay of charged
pions produced by cosmic ray interactions with interstellar gas,
primarily proton-proton interactions \cite{HIGHENERGYNEUTRINO}.
Some models to reproduce the observed ultra high-energy cosmic rays
also predicts ultra high-energy neutrinos as decay products
of superheavy dark matters \cite{EXTRADIMENSIONDM,RIGHTHANDEDNEUTRINO}.
Their flux is studied in \cite{NEUTRINOFROMSUPERHEAVYDM}.

\begin{figure}
\centerline{\psfig{file=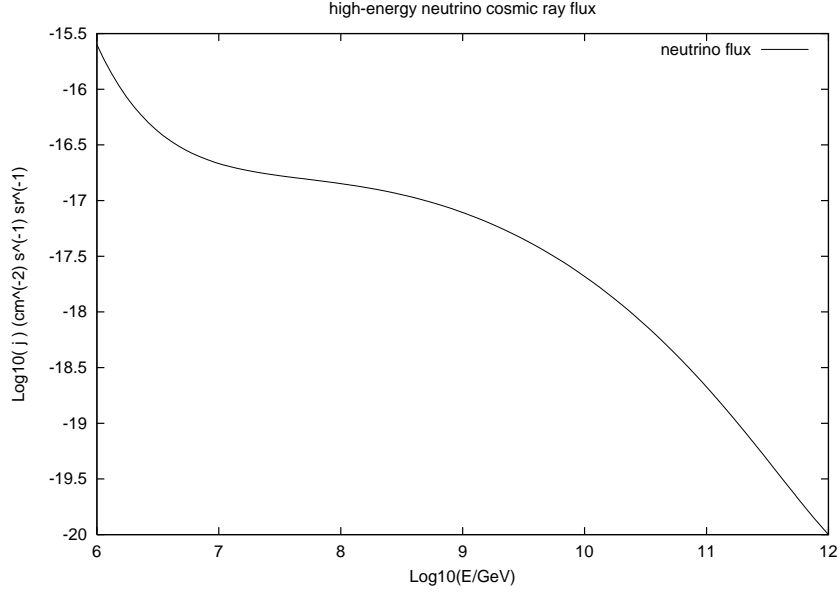,height=8cm}}
\caption{The expected high-energy neutrino cosmic ray flux.}
\label{NEUTRINOFLUX}
\end{figure}

The expected flux of high-energy neutrino was estimated in
\cite{HIGHENERGYNEUTRINO}, and it is drawn in figure \ref{NEUTRINOFLUX}.
The black hole production cross section by neutrino-nucleus scattering
is \cite{BHCR}
\begin{eqnarray}
\sigma(\nu N \rightarrow BH) = \sum_{i} \int_{(M_{BH}^{min})^{2}/s}^{1} dx \sigma_{i} (x s) f_{i} (x,Q).
\end{eqnarray}
Here, we have used MRST2001 
parton distribution function \cite{MRST}. 

Let $j(E)$ denotes the flux of high-energy neutrino cosmic ray and
$N$ denotes the number of neucleon in the detector. For $n=2$ and
$M_{*} = 1 \TEV$, the expected rate of black hole production $R$ becomes:
\bsub
\begin{eqnarray}
R &=& 4 \pi n \int_{M_{BH}^{min}}^{E_{\max}} \frac{d \sigma}{dE} j(E) dE \\
&=& 1.1 N \times 10^{-42} \ ({\rm events/sec}) = 3.5 N \times 10^{-35} \ ({\rm events/yr}).
\end{eqnarray}
\esub

Huge neutrino array like ICECUBE is the most suitable detector
for black hole. ICECUBE uses the ice of South Pole and its volume
is $1 {\rm km}^{3}$. It means $N=6 \times 10^{38}$. Thus ICECUBE can
produce about $21000$ black holes per year if $M_{*}=1 \TEV$ and $n=2$.
The dependence of the number of produced black holes on the fundamental
scale $M_{*}$ is shown in figure \ref{BHNUMBERFIG}.

\begin{figure}
\centerline{\psfig{file=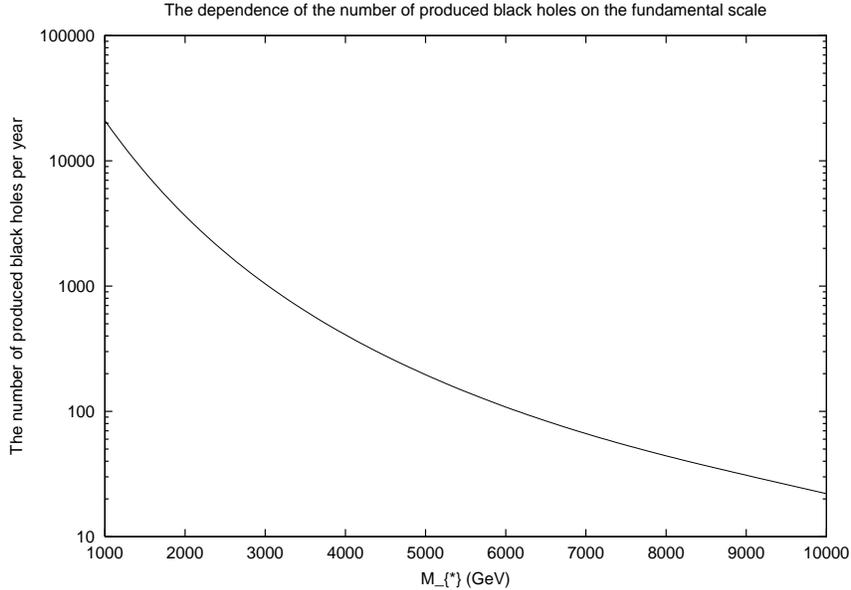,height=8cm}}
\caption{The number of produced black holes vs fundamental scale
 $M_{*}$. \label{BHNUMBERFIG}}
\label{BHNUMBERFIG}
\end{figure}

The main target of ICECUBE detector
is high-energy neutrinos whose energy is about ${\rm O}(\TEV)$.
High energy neutrinos coming up through the earth will 
occasionally interact with ice or rock and create a muon; 
such a muon emits Cherenkov light when passing through the array, 
and it can be tracked by measuring the arrival 
times of these Cherenkov photons at the PMTs.

Once produced, black holes immediately decays into the SM particles.
I do not enter its theory in detail. (see \cite{BHLHC}). 
The important point is that the decay of black
holes does not discriminate any particles: It decays into all particles
with roughly equal probability. The Standard Model contains about
$60$ particles, with 6 charged leptons. 
Black holes with mass ${\rm O}(\TEV)$ are mainly produced in ICECUBE detector,
As we can infer from the high-energy neutrino flux (figure \ref{NEUTRINOFLUX}).
Therefore we can observe the events of black holes from its decay into
hard leptons with energy ${\rm O}(\TEV)$.

To summarize, in this paper we consider the possible production 
of black holes caused by the collision between 
high-energy neutrino cosmic rays and nuclei in 
a large experimental detector. We observe that
ICECUBE detector should observe about $10^{4} \sim 10^{2}$ 
black holes per year if the fundamental 
scale is ${\rm O}(\TEV)$ and the number of
extra dimension is $n=2$. \footnote{Although there are severe
cosmological constraints on the fundamental scale and the number
of extra dimension \cite{SNCONSTRAINT,KKCONSTRAINT,INFLATIONCONSTRAINT}}

\noindent
{\bf Note added}

After we finished this paper, we learned from B.Harms that
they studied the evapolation of black holes in the presence 
of extra dimension \cite{EXTRADIMBH1,EXTRADIMBH2,EXTRADIMBH3}.
Their conclusion was that black holes in the presence of
extra dimensions would likely not evaporate.

\noindent
{\bf Acknowledgment}

We thank Y.Takeuchi and T.Yanagida for useful discussions.

\end{document}